\documentclass[a4paper, aps, prb, twocolumn,floatfix,  notitlepage, footnoteinbib]{revtex4-2}

\usepackage[T1]{fontenc}
\usepackage{graphicx}
\usepackage{amssymb}   % for math
\usepackage{amsfonts}
\usepackage{amsmath}
\usepackage{subfigure}
\usepackage{bm}        % for math
\usepackage[mathscr]{eucal}
\usepackage[dvipsnames]{xcolor}
\usepackage[colorlinks, linkcolor=blue,
citecolor=blue,
urlcolor=blue]{hyperref}

\usepackage[all]{hypcap} % let hyperlinks correctly point to figures rather than their captions;
\usepackage{physics}

\newcommand{\Le}{\left}
\newcommand{\Ri}{\right}

\newcommand{\f}{\frac}

\newcommand{\mc}{\mathcal}

\newcommand{\ra}{\rangle}
\newcommand{\la}{\langle}

\newcommand{\eq}[1]{\begin{align}#1\end{align}}

\newcommand{\msr}{\mathscr}

\newcommand{\Nc}{N_\text{csh}}
\newcommand{\Nsh}{N_\text{sh}}
\newcommand{\psh}{p_\text{sh}}

\newcommand{\Gc}{\Gamma_\text{c}}
\newcommand{\oNcNsh}{\overline{N_\text{csh}/N_\text{sh}}}
\newcommand{\betaL}{\beta_L(y, \Gamma)}
\newcommand{\Lx}{L_{\times}}
\newcommand{\Dgeo}{D_{\text{geom}}}

\begin{document}

\title{Finite-size geometric crossover toward fragmentation of the potential energy shell in the quantum random energy model}

\author{Ishita Modak}
\affiliation{Department of Physics, Indian Institute of Technology Bombay, Mumbai 400076, India}

\author{Vatsana Tiwari}
\affiliation{Department of Physics, Indian Institute of Technology Bombay, Mumbai 400076, India}

\author{Soumya Bera}
\affiliation{Department of Physics, Indian Institute of Technology Bombay, Mumbai 400076, India}

\date{\today}
\begin{abstract}
The quantum random energy model (QREM) is a paradigmatic model for studying delocalization-localization transitions at finite energy density in Fock space.
In the computational basis, its diagonal energies are independent Gaussian random variables, in contrast to the correlated energies of local interacting models such as the disordered XXZ chain.
In this work, we construct a potential energy shell associated with a basis state, defined through its energy variance, and investigate its connectivity on the Fock-space hypercube.
As the transverse field is varied, this energy-selected subgraph undergoes a
finite-size geometric crossover between fully connected and fragmented regimes.
Furthermore, we observe a correlation between the shell connectivity and the finite-size fractal dimension of the eigenstates, as extracted from the participation entropy.
For the QREM, the statistical independence of the diagonal energies allows the shell connectivity to be analyzed as an effective independent site percolation problem on the hypercube, with its crossover scale estimated within a one-hop approximation.
At finite energy density, this estimate gives a nonmonotonic dependence of the crossover scale on system size $L$: increasing $L$ initially favors connectivity, while beyond an energy-dependent scale $L^*(\varepsilon)$ the trend reverses towards fragmentation.

\end{abstract}
% \today{}
\maketitle

\section{Introduction}
Many-body localization (MBL) concerns the fate of Anderson localization in the presence of interactions at finite energy density~\cite{Basko2006,Gornyi2005,AletReview2018,SierantReview_2025}.
In contrast to one-dimensional non-interacting Anderson localization, which is well understood analytically, the MBL transition remains far from being fully understood, both analytically and numerically. 
Nonetheless, for a one-dimensional weakly interacting disordered spin chain with random local interactions, Imbrie rigorously established many-body localization, subject to an assumption limiting level attraction in the many-body spectrum~\cite{Imbrie2016, Imbrie2016-2}.
More recently, De Roeck \textit{et al.} proved that, in strongly disordered interacting quantum chains, the conductance of a chain of length $L$ vanishes faster than $1/L$, establishing the absence of normal heat conduction~\cite{DeRoeckProsniak25}.

Numerical studies of the localization transition, however,  remain severely limited by the exponential growth of the Fock-space dimension, which complicates the interpretation of the transition~\cite{Bera2017, Weiner19,PandaMBL19, NandyPRB21, AbaninAOP21,SierantPRB22,LongPRL23}.
More recently, many-body resonances have emerged as an important ingredient in the finite-size MBL crossover. Rare long-range resonances can persist in regimes where typical observables remain localized ~\cite{Morningstar2022,  ColboisPRB24,LaflorencieCatStatePRB25}, while their increasing presence or overlap at larger system sizes has been proposed as a route towards  delocalization.~\cite{CrowleySciP22,BiroliPRB24}.

The MBL problem can be represented as a single-particle localization problem on the graph of many-body basis states in Fock space~\cite{Altshuler1997,BiroliFSPRB17,LoganWelshPRB19, TarziaPRB20,  MonterioMBL21,
RoyFSAnatomyPRB21, detomasiFSPRB21,SutradharPRB22,ThibaultPRB24}. 
Within one such representation, the many-body basis states correspond to the vertices of the graph, the diagonal terms of the Hamiltonian play the role of on-site energies, and the physical hopping terms in real space, or more generally any off-diagonal terms such as particle-creation terms, generate hopping between nearby vertices of the graph.
Such a Fock-space is a high-dimensional graph defined by the $\sigma_z$ basis states with a branching number $K$ that is extensive with system size $L$. 
Moreover, every node in that high-dimensional graph has energies that are highly correlated~\cite{RoyFSPRB20,ThibaultPRB24}. 
Solving such a localization transition exactly, therefore, is a considerable challenge.

To isolate the roles of graph connectivity and correlations between
onsite energies, it is useful to consider Anderson localization on a
random regular graph.
The Anderson model on a random regular graph with fixed
branching number $K$ and independent onsite disorder exhibits a
well-established localization transition~\cite{TikhonovRRG16,
GarciaPRL17, BeraRRG18, GarciaMataRRGPRR20, TikhonovAnn21}.
In the large-connectivity limit, the
critical disorder scales as
$W_{\rm c}\sim K\ln K$~\cite{HerrePRB23,SierantSci23}.

Another approach is to construct graphs in which $K$ is extensive, but the node energies are uncorrelated; this is the quantum random energy model (QREM)~\cite{Laumann_PRL14, FaoroAnnPhys19, BiroliQREMPRB21, Manai2023, ThibaultPRB24}. 
The model shows a finite-size crossover towards localization with crossover scale as $W_\text{c} {\sim} \sqrt{L} \ln L$~\cite{ScoquartPRB25,BalducciRG25} (and it is therefore absent in the thermodynamic limit) in the middle of the band, i.e., at $\varepsilon=0$, where $\varepsilon$ is the many-body energy density. 
At finite energy densities $\varepsilon \ne 0$, the model shows a delocalization-localization transition, in contrast to $\varepsilon=0$~\cite{BaldwinPRB2016}.

In the Fock-space representation, typical eigenstates in an ergodic delocalized phase extend approximately uniformly over the full graph. In a localized phase, their weight is concentrated on a subset that occupies a vanishing fraction of the graph, although the number of contributing basis states may still grow exponentially with system size~\cite{Bauer2013}.
In this work, we provide a complementary geometric perspective on the QREM by studying the connectivity and fragmentation of its potential energy shell in Fock space.
In particular, we probe the energetically selected volume in Fock-space associated with a given initial energy density and investigate the statistical properties of this accessible volume. 
Recent work on the XXZ spin chain with a random field~\cite{ModakFrag25} has revealed a geometrical phase transition in the many-body Fock-space.
This transition corresponds to the fragmentation of Fock-space, unlike fragmentation driven by symmetries or kinetic constraints~\cite{Moudgalya_2022}.
The potential energy shell associated with the initial state fragments with increasing disorder, giving rise to a scenario in which the initial state becomes energetically disconnected from the rest of the potential energy shell. 
Additionally, such geometric fragmentation shows a strong correlation with the transient quantum dynamics: a fully fragmented shell exhibits slow relaxation dynamics, as most of the time-dependent wavefunction amplitude, at least at transient times, remains confined within the fragmented subspace~\cite{ModakFrag25}.

Here, we show that the QREM at $\varepsilon = 0$ exhibits a fragmentation crossover.
The corresponding crossover scale is found to be proportional to $L$, in contrast to the quantum delocalization crossover $W_\text{c} \propto \sqrt{L} \ln L$. The latter therefore grows more slowly with system size than the geometric crossover.
This difference could well be related to the fact that the quantum delocalization crossover involves coherent quantum interference, whereas the geometric fragmentation crossover is determined solely by the underlying connectivity and does not account for interference.
At finite energy density, the geometric crossover scaling slows down considerably and depends on the energy density $\varepsilon$. 
For the finite system sizes accessible in this work, we observe evidence of a possible fragmentation crossover. 
In contrast, at finite energy density, the QREM is believed to exhibit a localization transition~\cite{BaldwinPRB2016, BalducciRG25}.

To further understand the fragmentation of the potential energy shell, we show that the problem, at least for $\varepsilon=0$, admits an effective site-percolation description on the hypercube. 
Within this picture, the $1/L$ scaling of the crossover scale follows directly from the site-percolation criterion on the hypercube together with the Gaussian distribution of the onsite energies.
At finite energy density, however, the crossover scale is exponentially modified by the Gaussian density of states. 
Under the assumptions discussed in Sec.~\ref{sec:perc}, it is determined by the competition between the algebraic $1/L$ scaling and the exponential enhancement arising from the density of states.

%%%
%%%
\begin{figure}[!htb]
    \centering
    \includegraphics[width=0.85\columnwidth]{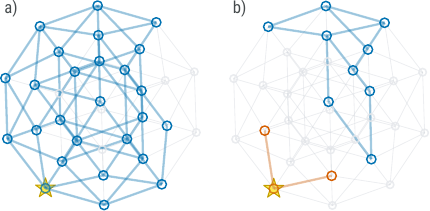}
    \caption{Illustration of the fragmentation of the potential energy shell $\msr{S}$ defined in Eq.~\eqref{eq:shell_define} for a basis state (shown here as a star) at energy density $\varepsilon=0$, and $L=5$ for a typical disorder configuration. The vertices denote basis states,  while the edges represent single-spin-flip connections as defined in Eq.~\eqref{eq:qremHam}. (a) For $\Gamma=0.6$, the shell forms a single connected component. (b) For $\Gamma=0.2$, the shell fragments into disconnected clusters (here two), confining the initial state to a subset of the energy shell. }
    \label{fig:qrem_demo}
\end{figure}
%%%
%%%

\section{Model and observables}
\subsection{Model: QREM}
In the computational basis (i.e., the $\sigma_z$ basis), the Hamiltonian for the quantum random energy model is defined as, 
\begin{equation}
    H = \Gamma \sum_i^L \sigma_i^x + \mathrm{diag} \{\epsilon_1, \epsilon_2, ... \epsilon_{2^L}\} 
    \label{eq:qremHam}
\end{equation}
where $\epsilon_i \   (i = 1, 2, \ldots, 2^L)$, are independent and identically distributed Gaussian random variables with zero mean and standard deviation $\sqrt{L/2}$, where $L$ is the system size and $\Gamma = 1/W$ denotes the transverse-field strength. On this basis, the QREM is equivalent to a single-particle hopping problem on an $L$-dimensional hypercube, where each vertex corresponds to a spin configuration, the transverse field induces hopping between nearest-neighbor vertices that differ by a single spin flip, and the Gaussian random energies $\epsilon_i$ act as on-site disorder.

At the band center, $\varepsilon=0$ (infinite temperature), the forward-scattering approximation predicts a finite-size localization crossover at $\Gamma_{\rm c}\sim(\sqrt{L}\ln L)^{-1}$ in the weak-field regime $\Gamma\ll1$.
While at finite energy density $\Gamma_\text{c}(\varepsilon)$ is finite in the thermodynamic limit~\cite{Laumann_PRL14, BaldwinPRB2016}. 
%

%%%%%%%%%%%%%%%%%%%%%%%%%%%%5
\begin{figure*}[!tbh]
    \centering
    \includegraphics[width=1\linewidth]{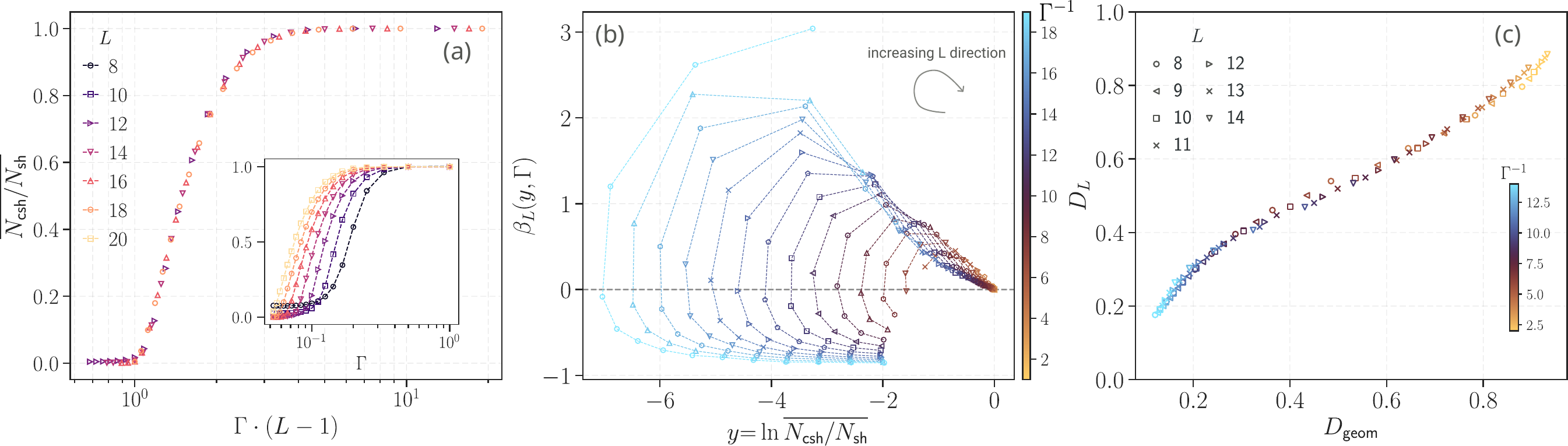}
    \caption{
    (a) Inset: Disorder dependence of the disorder-averaged connected fraction of the potential energy shell, $\overline{\Nc/\Nsh}$, for several system sizes $L$ at $\varepsilon=0$. Overline denotes the disorder average. The main panel shows an approximate data collapse when the horizontal axis is rescaled by $L-1$, indicating the finite-size crossover scale $\Gamma_{\mathrm{c}}\propto1/(L-1)$. For clarity, data for $L\lesssim12$ are omitted from the collapsed plot. (b) Corresponding $\beta$-function, $\betaL$ defined in Eq.~\eqref{eq:betafnc}, with $y=\ln \Le(\overline{\Nc/\Nsh}\Ri)$, showing the finite-size flow towards the fully connected regime. The lines serve as guides for the eye. (c) Correlation between the geometric dimension $\Dgeo$ of the connected shell and the finite-size fractal dimension $D_L$~\eqref{eq:entrp} for different $\Gamma$ (color scale) and $L$, showing an approximate collapse. The arrow in (b) indicates the direction of increasing system size $L$.
    }
    \label{fig:NcbyNshEzero}
\end{figure*}
%%%%%%%%%%%%%%%%%%%%%%%%%%%%5
\subsection{Geometric Observable}
Given a many-body configuration basis state $|\beta \ra$ with energy density $\varepsilon_\beta = E_\beta/L$, where $N = 2^L$ is the total number of basis states, the energy variance of the state is
\eq{
\Delta E = \Big[ \la \beta | H^2 | \beta \ra - \la \beta | H | \beta \ra^2 \Big]^{1/2}.
\label{eq:deltaE}
}
Using this energy width, we define a set
\eq{
\msr{S} = {i : |E_0 - E_i| < \Delta E },
\label{eq:shell_define}
}
where $E_0$ is the energy of the initial state.
Throughout this work, we refer to the set of basis states satisfying Eq.~\eqref{eq:shell_define} as the \emph{potential energy shell}, which is illustrated in Fig.~\ref{fig:qrem_demo}.

The resulting potential energy shell comprises the Fock-space basis states within the intrinsic energy uncertainty of the reference state. This choice is motivated by the fact that $\Delta E$ is the spectral width of the initial state, setting the energy window over which a typical wavefunction has appreciable support and, consequently, the characteristic short-time timescale $\tau \sim 1/\Delta E$ of decay of an observable~\cite{SantosEnergyShellPRE12}. The subsequent dynamics is therefore governed not only by the size of the potential energy shell but also by the connectivity of the set $\msr{S}$~\eqref{eq:shell_define}. In the following, we investigate the connectivity of this potential energy shell and its statistical properties. 

\paragraph*{Quantifying the Potential Energy Shell:} Throughout this work, we denote the number of basis states in the potential energy shell $\msr{S}$ by $\Nsh$. We denote by $\Nc$ the number of basis states in the connected cluster containing the initial state with energy density $\varepsilon$. Note that $\Nc$ does not necessarily correspond to the largest connected cluster within the potential energy shell.  

Fragmentation of the potential energy shell implies that the initial basis state becomes disconnected from a macroscopic fraction of the energetically accessible basis states, i.e., $\Nc/\Nsh \approx 0$. 
In contrast, when the potential energy shell is fully connected, $\Nc/\Nsh\approx1$, all energetically accessible basis states belong to a single connected component.

\subsection{Energy shell fragmentation in XXZ model}
The one-dimensional XXZ model with a random disorder field is a paradigmatic example of the crossover from the ergodic regime to the pre-MBL regime.
When mapped onto Fock-space, the model is described by a Hamiltonian similar to Eq.~\eqref{eq:qremHam}, except that the diagonal entries are correlated~\cite{RoyFSPRB20,  RoyFSAnatomyPRB21,  ThibaultPRB24,  ModakFrag25}.
Several works have identified an intermediate disorder regime, $W \gtrsim J, \Delta$, where $W$ is the disorder strength, $J$ is the XX and YY coupling, and $\Delta$ is the interaction strength, in which the thermalization dynamics is anomalously slow~\cite{LuitzPRL16, Weiner19}.

A recent work investigated these slow dynamics from the perspective of Fock-space geometry and the potential-energy shell~\cite{ModakFrag25}.
It was found that, at weak disorder, $W \simeq J, \Delta$, the sample-to-sample fluctuations of density imbalance are significant and are correlated with the fluctuations of the relative mass of the connected cluster, $\Nc/\Nsh$.
With increasing disorder, the potential energy shell fragments, and at a disorder strength $W_{\mathrm{frag}} \gg J, \Delta$, complete fragmentation of the shell occurs.
For finite system sizes, a crossover disorder strength, $W_{\mathrm{cross}}(L)$, can be defined, which increases logarithmically with system size, $W_{\mathrm{cross}}(L)\propto \log(L)$. 
Extrapolating this finite-size flow to a characteristic system size, $L^*\sim 10^2$, provides an estimate of the fragmentation disorder strength, $W_{\mathrm{frag}}$, at which the fragmentation transition occurs.
In this work, we precisely probe this crossover scale $W_\text{c}(L)$ (or equivalently $\Gamma_\text{c}(L)$) at different energy densities $\varepsilon$.

\section{Numerical results}
%%%%%%%%
\subsection{Finite size flow of $\Nc/\Nsh$ at $\varepsilon=0$}
%%%%%%%%
The inset of Fig.~\ref{fig:NcbyNshEzero}(a) shows the flow of disorder average connected cluster mass $\overline{\Nc/\Nsh}$ with $\Gamma$ for several system sizes $L$.
For $\Gamma \approx 0.5$, the hybridization induced by the off diagonal terms is strong, and the system exhibits a fully connected potential energy shell with $\overline{\Nc/\Nsh} \approx 1$.
As $\Gamma$ decreases below $\sim 0.25$, the potential energy shell begins to fragment for all $L$, and at even smaller $\Gamma$ it becomes completely fragmented, with $\overline{\Nc/\Nsh} \approx 0$.
The finite-size flow is shown in the main panel of Fig.~\ref{fig:NcbyNshEzero}(a), where the crossover point follows the scaling $\Gc  \propto 1/(L-1)$, as observed by an approximate collapse of the data over the entire range of system sizes when the x-axis is rescaled by $L-1$.

%%%%%%%%
\subsection{Finite size $\beta_L$ - function}
%%%%%%%
In the scaling theory of Anderson localization, the logarithmic flow of the conductance $g$ with system size $L$ in different dimensions determines the renormalization group (RG) structure of the localization transition~\cite{AbrahamsScalingTheoryPRL79}.
This quantity is described by the scaling function $\beta(g)=d\ln g/d\ln L$, which is universal and independent of microscopic details.

Recently, similar RG analyses have also been applied to the Anderson transition on high-dimensional graphs~\cite{AltshulerPNAS25,VanoniRRG24} and to the disordered XXZ model~\cite{NieddaRG2024}, using observables such as the energy-level spacing ratio and the eigenstate fractal dimension to characterize the finite-size flow of the critical disorder.
The $\beta$-function approach was also applied to the QREM by \textcite{BalducciRG25}, who reconstructed the flow of the gap ratio and the eigenstate fractal dimension at the energy densities $E=0$ and $E=\sqrt{\ln 2}/4$, and classified the resulting behavior in terms of one- and two-parameter scaling.
Motivated by these studies, here, we apply the $\beta$-function construction to a purely geometric observable, rather than a quantum observable: the connected fraction of the potential energy shell $\Nc/\Nsh$. Unlike the quantum observables considered previously, this quantity requires no diagonalization and, within the effective percolation picture~(see section \ref{sec:perc}), admits a closed-form finite-size prediction at $\Gamma \ll 1$.

%
%%%%%%%%%%%%%%%%%%
\begin{figure*}[!hbt]
    \centering
    \includegraphics[width=1\linewidth]{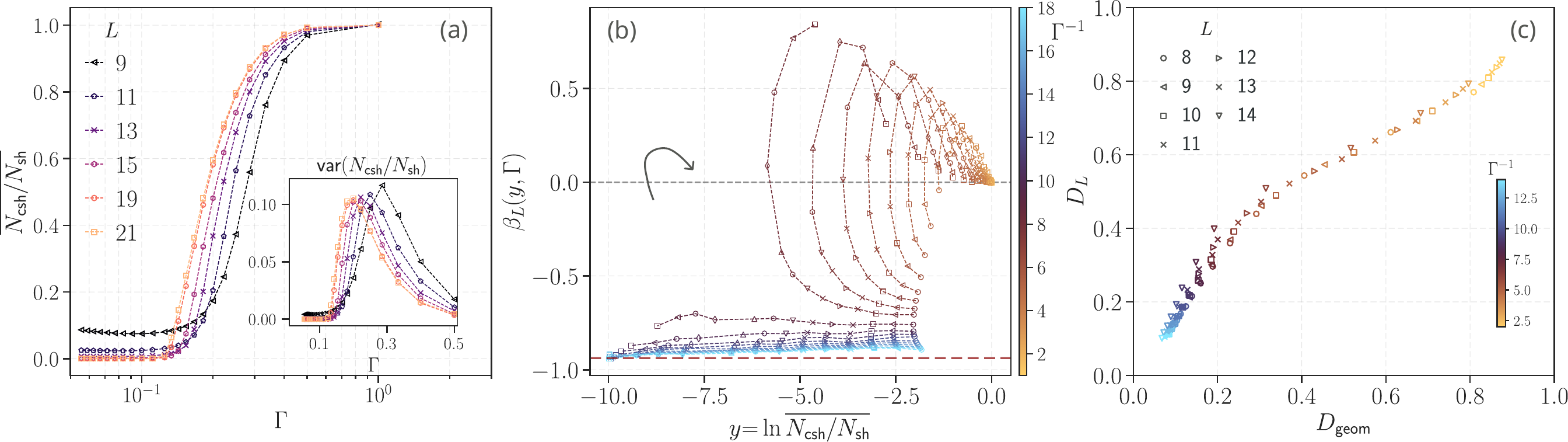}
    \caption{ (a) Disorder dependence of the disorder-averaged connected fraction of the potential energy shell, $\overline{\Nc/\Nsh}$, for various system sizes $L$ at $\varepsilon=\sqrt{\ln 2}/4$. Inset: Shows the $\text{var}(\Nc/\Nsh)$ as a function of $\Gamma$, highlighting the flow with $L$. (b) Corresponding finite-size flow of $x=\overline{\Nc/\Nsh}$ through the finite size function, $\betaL$, for various values of $\Gamma$. The horizontal dashed line denotes the asymptotic prediction $\beta_L^{\mathrm{sat.}}(\Gamma)\simeq-15/16$ from Eq.~\eqref{eq:betasat}. The lines serve as guides for the eye. (c) Correlation between $\Dgeo$ and the finite-size fractal dimension $D_L$ for various values of $\Gamma$ (color scale) and $L$. The arrow in (b) indicates the direction of increasing system size.}
    \label{fig:NcbyNshEfinite}
\end{figure*}
%%%%%%%%%%%%%%%%%
%
%

Motivated by these results, we here analyze the flow of $\Nc/\Nsh$ via its finite-size logarithmic derivative, 
\eq{
\beta_L(y, \Gamma)=\frac{d\ln x}{d\ln \mc{N}}; \qquad \mc{N}=2^L, 
\label{eq:betafnc}
}
here, $\mc{N}$ is the dimensionality of the Fock-space, and $y = \ln x= \ln (\overline{\Nc/\Nsh})$ where overline denotes the disorder average. 
We emphasize that this is a finite-size logarithmic derivative rather than an RG scaling function; we, however, retain the symbol $\beta_L$ by convention.
Since the QREM is naturally formulated in Fock-space, we use $\mc{N}$ as the scaling variable.

Figure~\ref{fig:NcbyNshEzero}(b) shows the $\beta_L(y, \Gamma)$ function for various values of $\Gamma$ and $L$.
%vasymptotic finite-size flow
As shown in panel (a), the asymptotic finite-size flow of $\overline{\Nc/\Nsh}$ is always towards the fully connected regime, that is, $\overline{\Nc/\Nsh}\approx 1$. Therefore, in terms of the logarithmic variable, this corresponds to the fixed point at $\beta_L(y^*=0,\Gamma)=0$.
Indeed, with increasing system size, the flow approaches the stationary point at $\beta_L(0,\Gamma)=0$, corresponding to the fully connected shell ($y^*=0$).
Close to this stationary point, the beta function is expected to be analytic and can therefore be expanded as
$
\betaL= -\lambda y + \mathcal{O}(y^2),  \, 
$
with $\lambda>0$.  

Since $\oNcNsh$, one has $x\le1$ and hence $y=\ln x\le0$, so
only the half-line $y\le0$ is physically accessible. This constraint fixes the
domain of the flow but not the sign of $\lambda$, which is instead read off from
the data. At large $\Gamma$, $\betaL(y)>0$ throughout the accessible range of $L$,
and the trajectories flow towards the stationary point at $y=0$, consistent with
$\lambda>0$. 

By contrast, at small $\Gamma$, $\betaL<0$ and the flow is directed away from the
stationary point towards complete fragmentation; at intermediate $\Gamma$ a single
trajectory exhibits both signs, $\betaL$ changing sign at a $\Gamma$-dependent
system size. Thus $\lambda>0$ characterizes the trajectories flowing into the
stationary point and is not a consequence of the constraint $x\le1$.

Now, 
$
y=\ln x=\ln\!\left[1-(1-x)\right]
=-(1-x)+\mathcal{O}\!\left((1-x)^2\right),
$
the $\betaL$ function may equivalently be written as
\eq{
\betaL=\lambda(1-x)+\mathcal{O}\!\left((1-x)^2\right).
\label{eq:beta_linear}
}
Thus, the leading approach to the stationary point is controlled by $\lambda$, while higher-order terms describe the nonlinear corrections to scaling.

Figure~\ref{fig:NcbyNshEzero}(b) also reveals the length scale, $\Lx(\Gamma)$, below which $\betaL <0$. Only for $L>\Lx (\Gamma)$ does $\betaL$ become positive, and the flow turns towards the fully connected regime.
We find that $\Lx(\Gamma) \propto \Gamma^{-1}$, consistent with the finite-size flow observed in Fig.~\ref{fig:NcbyNshEzero}(a) for $\Gc$, which is expected as $\betaL$ is a representation of the same data as shown in (a).

%%%%%%%%%%
\subsection{Correlation with fractal dimension $D_L$}
%%%%%%%%%%
We further probe the potential energy shell $\msr{S}$ through the participation entropy, defined as
\eq{
S_1(\varepsilon) = - \sum_j |\psi_j(\varepsilon)|^2 \ln |\psi_j(\varepsilon)|^2 
\sim D \ln \mc{N},
\label{eq:entrp}
}
where $D$ is the fractal dimension. Here, $D=1$ denotes a wavefunction delocalized over the Fock-space, while $D=0$ corresponds to a localized wavefunction.
$\psi_j(\varepsilon)$ denotes an eigenstate of Eq.~\ref{eq:qremHam}. 
We average over $10$ eigenstates around the energy density $\varepsilon$ and calculate the finite-size estimate of the fractal dimension, $D_L = \overline{S_1}/\ln \mc{N}$.

%\SB{
To compare the shell geometry with the eigenstate fractal dimension, we define
the geometric dimension of the connected potential-energy shell as
\eq{
\Dgeo=\frac{\overline{\ln \Nc}}{\ln\mc{N}}.
}
Figure~\ref{fig:NcbyNshEzero}(c) shows $D_L$ as a function of $\Dgeo$
for several values of $L$ and $\Gamma$. A few things are worth noting.
%
%

%\begin{itemize}
i) The data show an approximate collapse when plotted as a function of
$\Dgeo$ for all parameter values and system sizes. 

ii) Close to the fully connected regime $\oNcNsh\approx 1$, the agreement between $D_L$ and
$\Dgeo$ is expected, since the connected component occupies almost the
entire potential-energy shell, which at $\varepsilon=0$ forms an extensive
fraction of Fock space.

iii) The correlation persists as the connected component becomes smaller.
At small $\Gamma$, however, $D_L>\Dgeo$, showing that the eigenstates
cannot be confined to the connected shell component. 
The approximate collapse nevertheless suggests that the connected-shell volume provides a natural geometric scale for the finite-size support of the eigenstates.

iv) Nevertheless, $D_L$ and $\Dgeo$ probe different physics:
$D_L$ includes quantum interference, while $\Dgeo$ is purely geometric.
Their agreement should therefore be interpreted as a correlation, not as
evidence of causation.

%%%%%%%%%%%%%%%%%
\subsection{Finite energy density $\varepsilon= \sqrt{\ln 2}/4$}
%%%%%%%%%%%%%%%%%
\paragraph*{$\Nc/\Nsh$ flow:} At finite $\varepsilon$, the above picture changes considerably.
As shown in Fig.~\ref{fig:NcbyNshEfinite}(a), the flow of $\oNcNsh$ with $L$ slows down significantly as $\Gamma$ decreases.
In particular, for larger system sizes, $L\gtrsim16$, the residual flow is much weaker than the linear $L$ behavior observed at $\varepsilon=0$ in Fig.~\ref{fig:NcbyNshEzero}(a).
The inset shows the sample-to-sample variance of $\Nc/\Nsh$ as a function of $\Gamma$ for different system sizes. The peak marks the maximum fluctuations, and its shift as system size increases is consistent with the flow in the main panel.

\paragraph*{$\beta$-function:} The $\betaL$ reveals several features in the data, which we list below.
i) For $\Gamma \sim \mathcal{O}(1)$, the flow towards the stable stationary point at $\betaL=0$, corresponding to a fully connected potential energy shell, is clearly evident.

ii) In contrast to the $\varepsilon=0$ case, an extended intermediate-flow regime emerges at finite $\varepsilon$. Over a broad range of $\log(\oNcNsh)$, the $\betaL$ function evolves only weakly with increasing system size, indicating a slowing down of the finite-size flow. Within the accessible system sizes, the data are consistent
with the presence of an intermediate negative-$\beta$ plateau $\tilde{y}$, where $\beta(\tilde{y},\Gamma) < 0$.
Whether this corresponds to a genuine intermediate stable regime or merely a long finite-size crossover cannot be resolved from the present numerics.

iii) At smaller $\Gamma\lesssim0.1$, the flow is directed towards the fragmented regime. We observe a plateau extending over at least one decade in
$\overline{\Nc/\Nsh}$, whose saturation value is estimated by
Eq.~\eqref{eq:betasat}.

%, since $\psh=\erf(\Gamma)$ carries no $L$-dependence.

\paragraph*{Correlation with fractal dimension, $D_L$:} Figure~\ref{fig:NcbyNshEfinite}(c) shows the correlation between the
finite-size fractal dimension $D_L$ and the geometric dimension $\Dgeo$.
As in the $\varepsilon=0$ case, the data show an approximate collapse,
without any fitting parameter, over the range of parameters considered.
In the $D_L < 0.4 $ regime, the data also exhibit the qualitative finite-size
flow seen in the $\beta_L$ function, including a change in flow direction
with increasing $L$, consistent with a finite-size fragmentation crossover.

%%%
\section{Effective Site-Percolation Picture}
\label{sec:perc}
%%%
We analyze the geometric fragmentation crossover of the potential energy shell from a percolation perspective. We first evaluate the probability, $\psh$, that a many-body configuration (basis state) lies within the energy window $\Delta E$. The energy shell width is determined exactly from the local energy variance, $\Delta E=\Gamma\sqrt{L}$. 
Centering the window at the middle of the spectrum ($E=0$), this probability is
\eq{
p_\mathrm{sh}
=\mathrm{Pr}(|E|\le \Gamma\sqrt{L})
=\int_{-\Gamma\sqrt{L}}^{\Gamma\sqrt{L}}
\frac{e^{-E^{2}/L}}{\sqrt{\pi L}}\,dE
=\mathrm{erf}(\Gamma). 
\label{eq:psh}
}
The total number of configurations within the shell is therefore
$N_\mathrm{sh}=2^L p_\mathrm{sh}$.

To characterize the connectivity of the shell, we map the problem onto classical site percolation on an $L$-dimensional hypercube~\cite{Stauffer1994}. Each vertex representing a many-body configuration is connected to configurations differing by a single spin flip, and has a coordination number $L$. 
The mean number of shell configurations among the $L-1$ forward neighbours defines the branching ratio, 
$$
k_\mathrm{eff}
=(L-1) \cdot p_\mathrm{sh}
= (L-1) \cdot \mathrm{erf}(\Gamma). 
$$
Within the locally tree-like approximation (neglecting loops), the emergence of a connected cluster requires
$
(L-1)\cdot\mathrm{erf}(\Gc)=1,
$
which is the percolation threshold on an infinite Bethe lattice of coordination number $L$~\cite{Stauffer1994}.
For $\Gamma \ll 1$ in leading order,
$
\mathrm{erf}(\Gc)\simeq {2\Gc}/{\sqrt{\pi}},
$
yielding the following condition for a crossover scale, 
\eq{
\Gc \simeq \frac{\sqrt{\pi}}{2(L-1)}. 
\label{eq:gammac_L}
}
For $\Gamma<\Gc$, the shell consists of finite clusters with typical size $\Nc=O(1)$, implying $\Nc/\Nsh \rightarrow0$ as $L\rightarrow\infty$. In contrast, for any fixed $\Gamma>0$ independent of $L$, the effective connectivity diverges as $k_\mathrm{eff}\sim L$, and the shell contains a connected component with $\Nc/N_\mathrm{sh}\rightarrow1$.
Figure~\ref{fig:NcbyNshEzero}(a) indeed exhibits this linear scaling with $L$, consistent with the site-percolation threshold on the hypercube at $\varepsilon=0$. 
In contrast, for a graph with fixed connectivity, such as a random regular graph (RRG), the geometric crossover scale $\Gc$ is therefore expected to remain independent of $L$, and it is indeed consistent with the numerical results as seen in  Appendix~\ref{app:rrg}.

%%%%
\subsection{Finite energy density}
%%%%%%
At finite energy density, the shell probability $\psh$ defined in Eq.~\eqref{eq:psh} depends on $\varepsilon$ in the following way, 
\eq{
\psh(\varepsilon) =\f{1}{2}\Le[ \text{erf}(\varepsilon \sqrt{L}+\Gamma) - \text{erf}(\varepsilon \sqrt{L}-\Gamma) \Ri]. 
\label{eq:psh_E}
}
The above expression can be expanded in $\Gamma \ll 1$, and while factoring out the Gaussian term, gives, 
\eq{
\psh(\varepsilon)\simeq\frac{2\Gamma}{\sqrt{\pi}}\,e^{-\varepsilon^{2}L}
\Big[\,1+\tfrac16\lambda^{2}-\tfrac13\Gamma^{2}+\mathcal{O}(\Gamma^{4})\Big],
\label{eq:psh_expand}
}
where $\lambda= 2 \varepsilon \Gamma \sqrt{L} =2 \varepsilon \Delta E$.  
The term $e^{-\varepsilon^2 L}$ reflects the Gaussian suppression due to the density of states away from the band center. 
%
%
%%%%%%%%%%%%%%%%%%
\begin{figure}[!tbh]
    \centering
    \includegraphics[width=0.9\columnwidth]{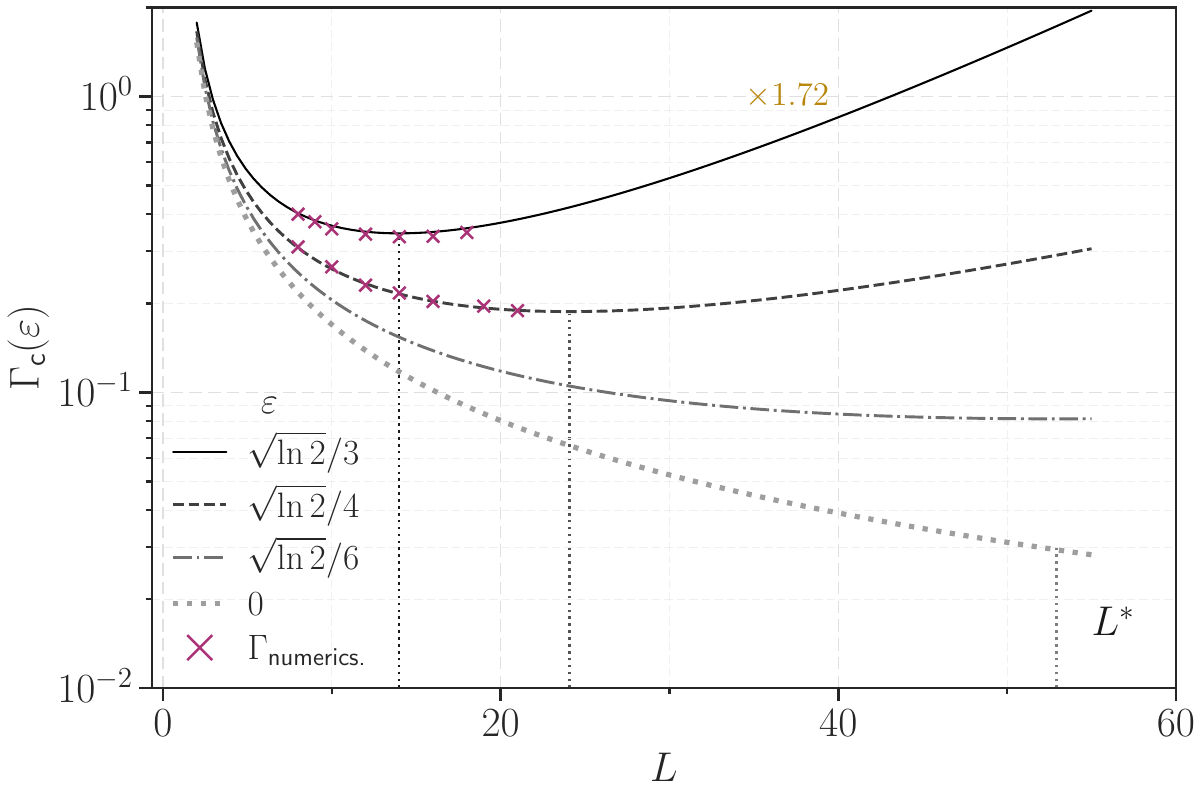}
    \caption{Finite-size one-hop estimate of the shell-percolation threshold, $\Gc(L,\varepsilon)$, at the indicated energy densities $\varepsilon$, as described in Eq.~\eqref{eq:gammac_E}.  A single heuristic multiplicative
factor of $\simeq 1.72$ is applied to the analytical curves to match the
numerical estimates, shown by crosses. For $\varepsilon\neq0$, $\Gc$ reaches a minimum at $L^*(\varepsilon)$ (vertical dotted lines): increasing $L$ initially
favors hypercube connectivity, but beyond $L^*(\varepsilon)$ disfavors it.
The numerical estimates, shown by crosses, are obtained from the maxima of $\text{var}(\Nc/\Nsh)$ 
(see Appendix~\ref{app:emax8}).}
    \label{fig:GammacL}
\end{figure}
%%%%%%%%%%%%%%%%%
%
%
To make further progress, we assume that the shell $\Delta E$ is sufficiently narrow that the Gaussian density of states varies little across it. This would require, 
$
{\rho(E+\Delta E)}/{\rho(E)}
=
\exp
\!\left[
-\lambda-\Gamma^2
\right] \approx 1, 
$
 hence $\lambda, \Gamma^2 \ll 1$. Retaining only the leading term of Eq.~\eqref{eq:psh_expand}, and using the locally tree-like approximation together with the hypercube percolation threshold $(L-1) \cdot p_\mathrm{sh}(\varepsilon) = 1$, we get the crossover scale as, 
\eq{
\Gc (\varepsilon) \simeq \f{\sqrt{\pi}}{2 (L-1)}{e^{\varepsilon^2L}}. 
\label{eq:gammac_E}
}
The exponential factor compensates for the exponentially decreasing density of states away from the band center and competes with the algebraic growth of the hypercube connectivity.
This expression is valid only while the narrow-shell condition $\lambda \ll 1 $ remains satisfied. 
Once $\lambda$ becomes of order unity, the narrow-shell approximation would fail, and the asymptotic solution crosses over to a different asymptotic branch.

\paragraph*{Finite-size crossover:} Figure~\ref{fig:NcbyNshEfinite}(a), however, shows a slowing down of the $1/L$ scaling, which may be related to the fact that the accessible system sizes are still in the pre-asymptotic regime.
Indeed, Eq.~\eqref{eq:gammac_E} predicts a minimum of $\Gc(L)$ at
$L^*=1+\varepsilon^{-2}$. For the energy densities considered here,
$\varepsilon=\sqrt{\ln 2}/4$ and $\varepsilon=\sqrt{\ln 2}/3$, this gives
$L^*\simeq 24$ and $L^*\simeq 14$, respectively, and is shown in Fig.~\ref{fig:GammacL} as vertical lines.
In the former case, the exponential and algebraic factors nearly cancel
across the accessible system sizes, accounting for the weak size dependence in
Fig.~\ref{fig:NcbyNshEfinite}(a); in the latter case, $L^*\simeq 14$ lies within the accessible system sizes,
and beyond this scale, the exponential suppression dominates over the
$1/(L-1)$ factor, leading to the turnaround towards fragmentation as seen in Fig.~\ref{fig:GammacL}.

The finite-size estimate Eq.~\eqref{eq:gammac_E} predicts a minimum at $L^*(\varepsilon)$, beyond which the density-of-states suppression dominates over the linear increase in hypercube connectivity. It is therefore natural to ask whether this finite-size connected regime survives asymptotically.
We evaluate the exact shell probability for any fixed energy density $\varepsilon>0$ in the limit $\varepsilon\sqrt{L}-\Gamma\gg1$. Using the leading asymptotic expansion of the error function, $\operatorname{erfc}(x)\approx e^{-x^2}/(x\sqrt{\pi})$, we obtain
\eq{
p_{\mathrm{sh}}
\approx
\frac{1}{2\sqrt{\pi}(\varepsilon\sqrt{L}-\Gamma)}
\exp\!\left[-(\varepsilon\sqrt{L}-\Gamma)^2\right].
\label{eq:psh_asymptotic}
}
Imposing the local tree-like percolation criterion $(L-1) \cdot p_{\mathrm{sh}}=1$ yields the transcendental equation
$
x e^{x^2}=(L-1)/(2\sqrt{\pi})$,
with $x=\varepsilon\sqrt{L}-\Gc$. Since $\ln x=\mathcal{O}(\ln\ln L)$ is subleading compared to $\ln L$, the asymptotic solution is $x\approx\sqrt{\ln L}$. Thus,
\eq{
\Gc(\varepsilon)
\approx
\varepsilon\sqrt{L}-\sqrt{\ln L}.
\label{eq:gammac_finiteE}
}
The narrow-shell approximation underlying Eq.~\eqref{eq:gammac_E}
requires $\Gamma^2\ll1$ and $2|\varepsilon|\Gamma\sqrt{L}\ll1$.
Since the asymptotic one-hop estimate in
Eq.~\eqref{eq:gammac_finiteE} grows as $\Gc\propto\sqrt{L}$, it lies outside this regime.
At fixed $\Gamma>0$ and $\varepsilon\neq0$, the exact shell
probability in Eq.~\eqref{eq:psh_E} satisfies
$Lp_{\mathrm{sh}}\rightarrow0$, so the reference configuration
becomes isolated within the shell with probability approaching one.
In other words, at nonzero energy density and fixed $\Gamma$, the connected regime is a finite-size effect, and increasing the system size ultimately leads to shell fragmentation.

\subsubsection{Saturation of $\betaL$ at $\Gamma \ll 1$}

The plateau at negative $\beta_L$ can be understood from the structure of Eq.~\eqref{eq:betafnc} along with Eq.~\eqref{eq:psh_expand}, which gives an estimate of the $\Nsh$ at finite $\varepsilon$. %
In this regime, where $\Nc \approx \text{const.}$ and $\Gamma\ll 1$~(or equivalently $W \gg 1$) the Eq.~\eqref{eq:betafnc} reduces to
\eq{
\beta_{L}^{\text{sat.}}(\Gamma)
\simeq
-1+\frac{\varepsilon^2}{\ln2}
- \f{2\varepsilon^2 \Gamma^2}{3 \ln 2} + \mathcal{O}(\Gamma^4). 
\label{eq:betasat}
}
For the energy density $\varepsilon=\sqrt{\ln 2}/4$ this gives the plateau values $\beta_L^{\rm sat}=-15/16$, and indicated in Fig.~\ref{fig:NcbyNshEfinite} as a dashed line. 
The correction is independent of $L$, so the plateau
is system-size independent within the present approximation. At the band centre ($\varepsilon=0$) the relation becomes independent of $\Gamma$, $\beta_L^{\rm sat}=-1$.

\section{Discussion and outlook}
In this work, we studied the geometric fragmentation crossover in the QREM across many-body energy densities.
We introduced the concept of the potential energy shell $\msr{S}$~\eqref{eq:shell_define} and investigated its connectivity. 
Our numerical results suggest a finite-size crossover between connected and fragmented shells at the band center ($\varepsilon=0$), while at finite energy density, the observed behavior remains pre-asymptotic. 
Within the effective mean field site-percolation description of the shell geometry, $\oNcNsh$ measures the fraction of shell sites connected to the initial configuration. 
In this picture, the crossover scale follows $1/(L-1)$ at $\varepsilon=0$, while at finite $\varepsilon$ it is governed by the competition between the algebraic $1/(L-1)$ connectivity of the hypercube and the exponential suppression of shell sites by the density of states, Eq.~\eqref{eq:gammac_E}.
Finally, we observed a strong correlation between the fractal dimension $D_L$ and the shell geometry, with an approximate collapse over all accessible $L$ and $\Gamma$, particularly for $\varepsilon=0$. 
Therefore,  the connected energy shell volume may serve as a geometric measure of finite-size eigenstate support.

% \SB{suggesting that the eigenstates predominantly occupy the connected component of the energy shell. }

We emphasize that the accessible numerical system sizes remain within the pre-asymptotic regime.
Since the present analysis in Sec~\ref{sec:perc} is based on single-hop resonances, Eqs.~\eqref{eq:gammac_E} should be viewed as a finite-size crossover estimate rather than a quantitative description of the current numerics.
It therefore remains unclear whether the observed finite-size flow eventually follows the predicted one-hop asymptotics or crosses over to a different regime.
Nonetheless, for $\varepsilon=\sqrt{\ln2}/3$, a turnaround in $\Gc(L)$ is
observed in Fig.~\ref{fig:GammacL}, which is consistent with the estimate
$L^*\simeq14$ from the one-hop condition. 
For the quantum problem, however, such an estimate would require corrections; in particular, one would need to incorporate higher-order, multi-hop processes and loops.

Several remarks are in order.
First, the fragmentation discussed here is energetic rather than an exact
fragmentation of Hilbert space. The full Fock-space hypercube remains
connected, while the energy-selected shell can break into disconnected
components. These components may still be connected through states outside
the shell.
Accordingly, it is not expected to capture either the quantum
delocalization--localization transition~\cite{Laumann_PRL14,BaldwinPRB2016}
or the finite-size scaling
$\Gamma_{\rm c}^{-1}\sim W_{\rm c}\sim\sqrt{L}\ln L$
reported in previous works~\cite{ScoquartPRB25,BalducciRG25}.
Furthermore, our analytical argument is based on a one-hop percolation
criterion and neglects the loop structure of the hypercube.

This separation between the geometric crossover scale and the quantum resonance scale is already visible at the bare single-hop level and can be understood from a simple level-spacing argument. At the band centre, the $L-1$ forward neighbours
of $|\beta\rangle$ have a characteristic spacing
$\delta_L=[(L-1)\rho(\varepsilon{=}0)]^{-1}\simeq\sqrt{\pi/L}$. For $\Gc\ll1$,
the shell width at the crossover,
$\Delta E=\Gc\sqrt{L}\simeq \frac{1}{2}\sqrt{\pi/L}$,
is therefore of order $\delta_L/2$. By contrast, a bare single-hop
resonance compares $\delta_L$ directly with $\Gamma$. Since
$\Delta E/\Gamma=\sqrt{L}$, the two criteria scale as
$\Gc^{\rm geom}\sim L^{-1}$ and
$\Gamma^{\rm res}\sim L^{-1/2}$.

Second, previously, Roy \textit{et al.} constructed classical Fock-space
percolation proxies in which edges between basis configurations are
retained when the corresponding states are expected to hybridize
strongly~\cite{RoyPercPRB19,RoyPercPRB19a}. A local resonance criterion is
$
    |E_i-E_j|<|V_{ij}|.
$
Within this framework, the many-body localized phase is associated with the fragmentation of the resonant network.
Prelov\v{s}ek \textit{et al.} instead represent the system in the
many-body basis of noninteracting Anderson-localized orbitals and,
after retaining resonant processes, reduce the dynamics to an
approximate rate-equation network~\cite{PrelovsekPercPRB21}. In both
constructions, localization is associated with the loss of a
macroscopically connected resonant cluster.
In contrast, our work employs a global criterion based on the potential-energy shell and examines its connectivity. The shell is defined by the energy variance of the initial basis state~\eqref{eq:shell_define} and provides a coarse-grained geometric description that would contain several local resonances. 
Therefore, it would be interesting in future work to compare the finite-size scaling emerging from such local resonance criteria with that obtained from the present global shell construction, and to understand whether the two approaches ultimately belong to the same scaling framework.

Finally, we note that in the present geometric approximation, Eqs.~\eqref{eq:psh} and \eqref{eq:psh_E}, the shell-fragmentation crossover occurs at disorder strengths that are parametrically larger than those of the known localization transition.
Whether this separation persists when quantum effects from higher-order resonance processes are included to form the $\Nc/\Nsh$ remains an open question.
A natural direction for future work is to incorporate multi-step resonances or develop a self-consistent construction of the resonance graph and investigate the fragmentation properties of the resulting graph.

We conclude by noting that recent advances in quantum simulation experiments now allow direct monitoring of Fock-space dynamics~\cite{YaoNat2023}. Although such experiments do not directly realize the QREM, it would be interesting to investigate whether the energy shell fragmentation identified here has broader implications for many-body dynamics.

\section{Acknowledgment}
We thank F Balducci and G De Tomasi for a critical reading of the manuscript and for several discussions. 
We also thank F Evers for collaboration in an earlier related work~\cite{ModakFrag25}, which motivated the current study. 
We also thank I Gornyi for several useful discussions.   

IM acknowledges financial support from the Prime Minister's Research Fellows (PMRF) scheme offered by the Ministry of Education, Government of India.
SB thanks the National Supercomputing Mission~(NSM) for providing computing resources of `PARAM Rudra' at IIT Bombay, implemented by C-DAC and supported by the Ministry of Electronics and Information Technology (MeitY) and
Department of Science and Technology~(DST), India.
The authors thank the Yukawa Institute for Theoretical Physics at Kyoto University. Discussions during the YITP workshop YITP-W-26-10 on `Localization26' are gratefully acknowledged. 
We used ChatGPT (GPT 5.6 Sol) to assist with checking derivations and for language editing.

\setcounter{figure}{0}
\renewcommand{\thefigure}{A\arabic{figure}}
\renewcommand{\figurename}{Figure}
\appendix

\section{Sample to sample fluctuations}
\label{app:emax8}

%%%%%%%%%%%%%%%%%%
\begin{figure}[!hbt]
    \centering
    \includegraphics[width=1\columnwidth]{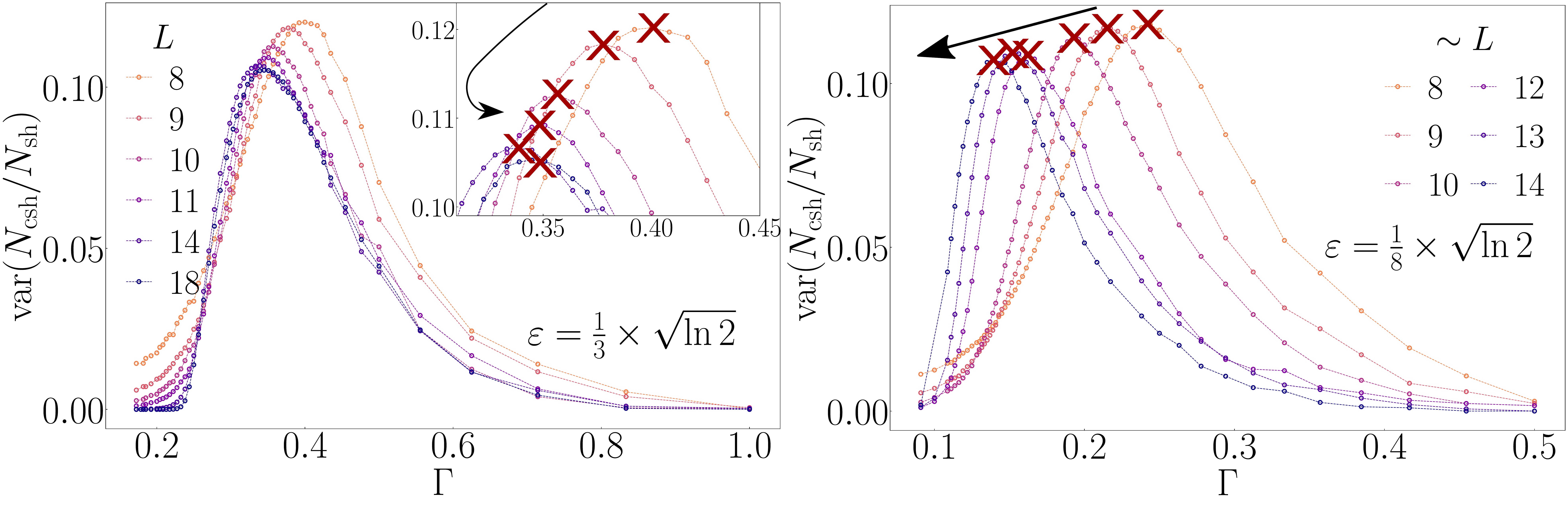}
    \caption{Sample-to-sample variance of the connected shell fraction,
$\Nc/\Nsh$, as a function of $\Gamma$ for several system sizes $L$ at
(a) $\varepsilon=\sqrt{\ln2}/3$ and
(b) $\varepsilon=\sqrt{\ln2}/8$. For each $L$, the finite-size
numerical crossover scale, $\Gc(L,\varepsilon)$, is defined as the position of the variance maximum. The red crosses mark these maxima in the inset of (a); the extracted values are the numerical estimates shown as crosses in Fig.~\ref{fig:GammacL}.} 
    \label{fig:var_0.33_0.125}
\end{figure}
%%%%%%%%%%%%%%%%%
For each $L$ and $\varepsilon$, we calculate the sample-to-sample
variance $\text{var}\left(\frac{\Nc}{\Nsh}\right)$. 
The maximum of this variance provides a finite-size estimator
$\Gc(L,\varepsilon)$ of the geometric crossover,
where the sample-to-sample fluctuations of the connected shell fraction are largest. For $\varepsilon=\sqrt{\ln2}/3$, the extracted $\Gc$ decreases with $L$ up to
$L\simeq14$ and subsequently turns upward, consistent with the
one-hop estimate $L^* \simeq14$ in
Fig.~\ref{fig:GammacL}. For $\varepsilon=\sqrt{\ln2}/8$, the extracted crossover scale decreases
monotonically throughout the accessible range. In this case, the
predicted turnaround occurs only at
$L^*\simeq94$ for the system sizes
used here, well beyond the numerically accessible range. The absence
of a visible turnaround in panel (b) is therefore consistent with the
one-hop finite-size estimate.

\section{Fragmentation on Random Regular Graph}
\label{app:rrg}
We study energy-shell fragmentation on a random regular graph (RRG)
with $N=2^L$ vertices and fixed degree $K$, independent of $L$. The
Hamiltonian is
\begin{equation}
    H =
    -t\sum_{\langle i,j\rangle\in\mathrm{RRG}}
    \left(
        \ket{i}\bra{j}+\ket{j}\bra{i}
    \right)
    +
    \operatorname{diag}
    \{\epsilon_1,\epsilon_2,\ldots,\epsilon_{2^L}\},
    \label{eq:rrgHam}
\end{equation}
where $\langle i,j\rangle$ denotes an edge of the graph. Each vertex
has degree $K$, and we use $K=3$ throughout. The diagonal energies are
independent random variables drawn from
$
    \epsilon_i\sim
    \mathcal{U}\left[
        -\frac{1}{2\Gamma},
        \frac{1}{2\Gamma}
    \right],
$
where $W\equiv1/\Gamma$ is the disorder strength. For $K=3$ and $t=1$,
the Anderson localization transition occurs at
$W_{\rm c} \simeq18$, corresponding to
$\Gamma_{\rm c}\simeq1/18$~\cite{TikhonovRRG16,BeraRRG18}.

\begin{figure}[!thb]
    \centering
    \includegraphics[width=0.7\columnwidth]
    {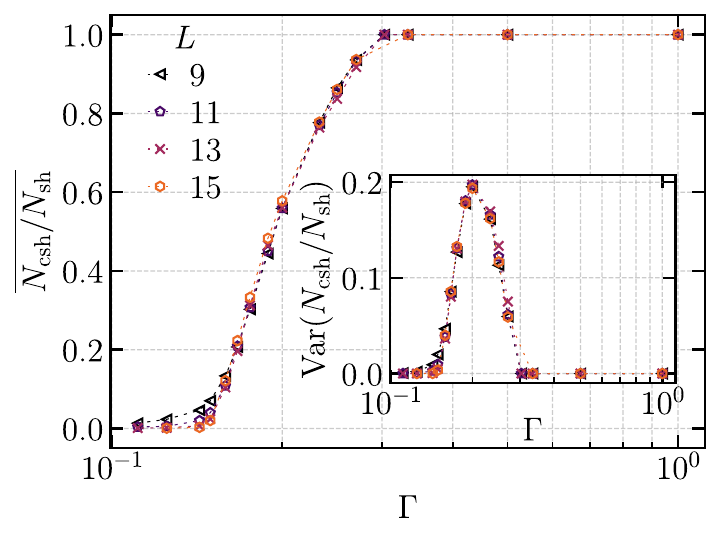}
    \caption{
    Disorder-averaged connected fraction of the energy shell,
    $\overline{N_{\mathrm{csh}}/N_{\mathrm{sh}}}$, as a function of
    $\Gamma$ for system sizes $L=8$--$15$ on an RRG with fixed degree
    $K=3$. The weak system-size dependence is consistent with the
    size-independent one-hop crossover scale in
    Eq.~\eqref{eq:rrg_gammac}.
    }
    \label{fig:rrg_fragmentation}
\end{figure}
Using the general definition in Eq.~\eqref{eq:deltaE}, the energy
fluctuation of a basis state $\ket{i}$ is
$
    \Delta E_i
    =
    \sqrt{
        \langle i|H^2|i\rangle
        -
        \langle i|H|i\rangle^2
    }
    =
    t\sqrt{K}.
    \label{eq:rrg_deltaE}
$
According to Eq.~\eqref{eq:shell_define}, a neighbouring vertex $j$
belongs to the energy shell of $\ket{i}$ when
$
    |\epsilon_j-\epsilon_i|\leq\Delta E_i.
$

For a shell centered at the band center, the probability that a vertex
belongs to the shell is
\begin{equation}
    p_{\rm sh}
    =
    \int_{-\Delta E}^{\Delta E}\Gamma\,dE
    =
    2t\sqrt{K}\,\Gamma,
    \label{eq:rrg_psh}
\end{equation}
provided that the shell interval lies within the support of the
disorder distribution,
$\Delta E\leq1/(2\Gamma)$. When the shell covers the entire
distribution, this probability instead saturates at $p_{\rm sh}=1$.

Applying the locally tree-like one-hop criterion introduced in
Sec.~\ref{sec:perc}, with the hypercube branching number $L-1$
replaced by the RRG branching number $K-1$, the effective branching
ratio is
$
    k_{\rm eff}
    =
    (K-1)p_{\rm sh}.
$
The site-percolation threshold is determined by
$k_{\rm eff}=1$~\cite{Stauffer1994}, which gives
\begin{equation}
    \Gamma_{\rm c}^{\rm geom}
    =
    \frac{1}{2t\sqrt{K}\,(K-1)}.
    \label{eq:rrg_gammac}
\end{equation}
For $K=3$ and $t=1$,
$
    \Gamma_{\rm c}^{\rm geom}
    =
    \frac{1}{4\sqrt{3}}
    \simeq0.144.
$
Because $K$ is independent of $L$, the one-hop estimate therefore
predicts a size-independent geometric crossover scale.

Figure~\ref{fig:rrg_fragmentation} shows that the finite-size flow of
$\oNcNsh$ depends only weakly on $L$, in contrast to the QREM, whose
degree grows as $L$. This behavior is consistent with the
size-independent geometric crossover predicted by
Eq.~\eqref{eq:rrg_gammac}. The geometric threshold
$\Gamma_{\rm c}^{\rm geom}\simeq0.144$ should not be identified with
the Anderson localization threshold
$\Gamma_{\rm c}^{\rm And}\simeq1/18$: the former characterizes
connectivity of the energy-selected subgraph, whereas the latter is a
quantum localization transition.

\bibliography{MBL}

\end{document}